\documentclass[twocolumn,showpacs,preprintnumbers,amsmath,amssymb,superscriptaddress]{revtex4-1}
\usepackage{graphicx}% Include figure files
\usepackage{dcolumn}% Align table columns on decimal point
\usepackage{bm}% bold math

\begin{document}

\title{Past and future blurring at fundamental length scale}

\author{M.J. Neves}
\affiliation{Centro Brasileiro de Pesquisas F\'\i sicas, Rua Dr.
Xavier Sigaud 150, Urca, Rio de Janeiro, RJ, 22290-180, Brasil}
\author{C. Farina}
\author{M.V. Cougo-Pinto}
\affiliation{Instituto de Fisica, UFRJ, CP 68528, Rio de Janeiro,
RJ, 21941-972, Brasil}

\begin{abstract}
We obtain the $\kappa$-deformed versions of the retarded and
advanced Green functions and show that their causality properties
are blurred in a time interval of the order of a length parameter  $q=1/(2\kappa)$. The
functions also indicate a smearing of the light cone. These results
favor the interpretation of $q$ as a fundamental length
scale below which the concept of a point in spacetime should be
substituted by the concept of a fuzzy region of radius $q$, as
proposed long ago by Heisenberg.
\end{abstract}

\maketitle

In the early days of quantum field theory (QFT) Heisenberg analyzed
several arguments for the introduction of a universal invariant
length parameter in physics \cite{Heisenberg38}, say $q$, which
should be added to the fundamental speed $c$ and the fundamental
action $\hbar$. One of Heisenberg's arguments relies upon March's
\cite{March36-37} proposal that geometry should be modified at small
lengths. Heisenberg argues that this length establishes an upper
limit in the accuracy of measuring positions and that the concept of
length itself can be used without restriction only for distances
which are large compared to the fundamental length.
The invariant character of such fundamental length follows then from
the relativistic principle \cite{Amelino-Camelia2001-2002}.
It is well known that the combination of gravitational and quantum effects implies
indeed the existence of such a length
\cite{DoplicherFredenhagenRoberts1994}, usually
at Planck scale, {\it i.e.}, of
order $10^{-33}$ cm, and several recent proposals have supported the
idea of introducing such a fundamental length (or the
corresponding fundamental energy) (see, {\it e.g.}\cite{Amelino-Camelia2001-2002,DSR}
and the reviews \cite{LIV}). A fundamental length of the order of $10^{-32}$
cm has also been advocated as a consequence of string theory (for a
review, see \cite{Witten96}), in a way which suggests that below
such a length \lq\lq{the smaller distances are not there}\rq\rq, to
use Witten's words \cite{Witten96}. In this context it is natural to
substitute the concept of a point in space by a fuzzy region of
radius $q$ and the concept of an instant of time by a fuzzy time
interval of duration $q/c$. More generally, a point in spacetime
would be substituted by a fuzzy region of radius $q$, if we used
natural units, as we do from now on. A quite remarkable implication
of such a view is a relaxation of the concept of causality, since a
portion of immediate future or past time of the order of $q$, for
example, would be indistinguishable from the present time. To face
this rather far from intuitive concept we may cite
Heisenberg's remark that the fundamental length together with the
constants $c$ and $\hbar$ designate \lq\lq{the limits in whose
proximity our intuitive concepts can no longer be used without
misgivings}\rq\rq \cite{Heisenberg38}.
A way of introducing a fundamental length $q$ at which our intuitive
concept of spacetime point becomes fuzzy is to implement a
non-commutative geometry of spacetime
\cite{Szabo2003-DouglasNekrasov2001} by promoting the spacetime
coordinates to hermitian operators obeying relations of the form
$[x^\mu,x^\nu]=iq^2\theta^{\mu\nu}$, in which we take
$\theta^{\mu\nu}$ dimensionless to make explicit the presence of
$q^2$ in the commutation relations. In the original formulation of
Snyder \cite{Snyder1947} $q^2\theta^{\mu\nu}$ is postulated to be an
operator of the Lorentz algebra, but more recently the
non-commutativity of space coordinates has been obtained as an
appropriate limit of string theory in non-trivial backgrounds
\cite{ConnesDouglasSchwars1998-DouglasHull1998-SeibergWitten1999}.
These non-commutative theories present nonlocality in spatial
directions which ruins Lorentz invariance and in the case of
non-commutativity of space coordinates and time acausal effects may
also appear \cite{SeibergSusskindToumbas2000,ChuFurutaInami2006}.
Actually, it can be argued on general grounds that violation of
Lorentz invariance or locality lead to violations of unitarity or
causality \cite{Campo2010}. Another way of introducing a fundamental
length $q$ is to deform the usual Lorentz invariant dispersion
relation in usual spacetime \cite{LIV}. We will consider here such a
deformation to investigate its effect on causality properties.
It is certainly worthwhile to investigate
if a theory containing a fundamental length $q$ shows
plausible and expected properties or unavoidable inconsistencies.
For instance, we could check if such a theory exhibits the above mentioned
{\lq\lq}relaxed causality{\rq\rq} due to the existence of a
fundamental length $q$. The appropriate quantities to investigate
these matters of causality are the Green functions for signal
propagation, namely, in the usual case, the retarded and advanced
Green functions of the wave operator $(\partial/\partial
t)^2-\nabla^2$. In this letter, we consider a theory in which such
operator is substituted by the deformed operator depending on a
length parameter $q$, given by $q^{-2}\sin^2(q\partial/\partial
t)-\nabla^2$, whose origin will be explained below. Note that this
operator reduces to the usual wave operator in the limit
$q\rightarrow 0$. A Green function associated to this deformed
operator satisfies the differential equation
\begin{eqnarray}\label{equacaogreen}
\left[\!\frac{1}{q^2}\sin^2\!\!\left(\!q\frac{\partial}{\partial
t}\!\right)
 \!\! - \!\!\nabla^{2}\right]\!\! G({\bf x},t;{\bf
 x}^{\prime},t^{\prime})\!=\!
-\delta^3({\bf x}\!-\!{\bf x}^{\prime})\delta(t\!-\!t^{\prime}).\;\;\;
\end{eqnarray}
At first sight there is no reason for this Green function to exhibit
the above mentioned property of blurring the concepts of past and
future inside time intervals of order $q$. In other words, there is
no {\it a priori} reason for the parameter $q$ in
(\ref{equacaogreen}) to play the role of a fundamental length. The
motivation to investigate this possibility is given by the origin of
this particular deformed operator, to which we pass now.

Due to the assumed smallness of the fundamental length $q$, a theory
describing phenomena at scale $q$ should be at least a theory of
quantum fields, if not of more fundamental objects. A natural place
to introduce a fundamental length in QFT is in its algebra of
space-time symmetries, since the fundamental length $q$ is obviously
a property of space-time. In usual relativistic QFT the algebra of
space-time symmetries is the Poincar\'e Lie algebra, defined by the
commutation relations of the generators $P^0$ and ${\bf P}$ of time
and space translations, as well as the generators of space rotations
and boosts. A fundamental property of this algebra is that ${\bf
P}^2-P_0^2$ is a Casimir invariant, {\it i.e.}, it commutes with all
the generators of the algebra. From this invariant we obtain the
wave operator by the usual substitutions $P^0\mapsto
i\partial/\partial t$ and ${\bf P}\mapsto -i\nabla$. The
introduction of a length $q$ in this algebra can be done by a
deformation of the algebra. The mathematical procedure of deforming
an algebra  consists in defining a new algebra depending on the
parameter $q$ with the property that the original one is recovered
in the limit $q\rightarrow 0$ of no deformation. In this context $q$
is called the deformation parameter. The deformation of the Poincar\'e
Lie algebra can be done in such a way that the deformed algebra is
part of a mathematical structure called Hopf algebra, also known as
quantum group or quantum algebra (see, {\it e.g.}
\cite{ChaichianDemishev96,Majid95}). This is a mathematical object
used to describe generalized symmetries not properly described by
Lie algebras \cite{ChaichianDemishev96}. A deformation of this kind
has been obtained by Lukierski, Ruegg, Nowicki and Tolstoy
\cite{LukierskiRueggNowickiTolstoy91,LukierskiNowickiRuegg92} and is
called $\kappa$-deformed Poincar\'e algebra, from the $\kappa$ used to
denote the deformation parameter with dimension of mass. Among the
remarkable properties of the $\kappa$-deformation of the Poincar\'e
algebra one is especially important here, namely: it is the simplest
deformation of the Poincar\'e algebra which gives rise to a quantum
group which provides a theoretical laboratory for QFT investigations
\cite{MajidRuegg94}. The $\kappa$-deformation of the Poincar\'e
algebra is defined by commutation relations between generators $P^0$
and ${\bf P}$ of time and space translations, as well as the
generators of space rotations and boosts
\cite{LukierskiRueggNowickiTolstoy91,LukierskiNowickiRuegg92} and
has the Casimir invariant ${\bf P}^2-(2\kappa)^2\sinh^2[P_0/(2\kappa)]$.
Using in this expression the usual substitutions $P^0\mapsto
i\partial/\partial t$ and ${\bf P}\mapsto -i\nabla$ we obtain the
$\kappa$-deformed wave operator
$(2\kappa)^2\sin^2[(2\kappa)^{-1}\partial/\partial t]-\nabla^2$. The
trivial substitution in this operator of the parameter $\kappa$ by
the length parameter $q=(2\kappa)^{-1}$ leads to the deformed
operator in (\ref{equacaogreen}). Regardless of these motivations
for adopting  here this deformed operator, we should observe that
all the calculations and investigation of causality questions of
this letter follow from the sole assumption of equation
(\ref{equacaogreen}). Let us, then,  calculate the $\kappa$-deformed
retarded and advanced Green functions. We start by considering the
following Fourier transform of the Green function in
(\ref{equacaogreen}):
\begin{eqnarray}\label{integralfourierespacialtemporal}
G^{(\gamma)}({\bf x},t;{\bf x}^{\prime},t^{\prime}) =
 \int\!\!\frac{d^3{\bf k}}{(2\pi)^3} \; e^{i{\bf k}.({\bf x}-{\bf
x^{\prime}})}\; {\cal G}^{(\gamma)}({\bf k},t-t^{\prime}) \; ,
\end{eqnarray}
where
\begin{eqnarray}\label{IntegraltemporalICalC}
{\cal G}^{(\gamma)}({\bf k},t-t^{\prime})=\int_{\gamma}
\frac{d\omega }{2\pi} \frac{e^{-i\omega
(t-t^{\prime})}}{q^{-2}\sinh^{2}(q\omega )-{\bf k}^2} \; ,
\end{eqnarray}
and $\gamma$ is a contour in the complex $\omega  $-plane infinitesimally close to the real axis
but bypassing the singularities in this axis. The poles on the complex $\omega  $-plane are given
by
$ \omega_{n}=\pm\omega_{\bf k}+ {2\pi i n}/{q}$ ($n\in {\mathbb Z}$),
where $\omega_{\bf k}=q^{-1}\sinh^{-1}(q|{\bf k}|)$. The choice of
 $\gamma$ bypassing the poles at points $\omega
=\pm\omega_{\bf k}$ on the real axis is equivalent to displace them
by infinitesimal imaginary quantities. The displacement
$\pm\omega_{\bf k}\mapsto \pm\omega_{\bf k}-i\varepsilon$ gives
rise to the retarded Green function of the wave equation and the
displacement $\pm\omega_{\bf k}\mapsto \pm\omega_{\bf k}+i\varepsilon$, 
to the advanced Green function of the wave
equation. These Green functions can be
calculated by the usual complex variable  technique of closing the
contour $\gamma$ by an infinite semicircle on the upper or lower
half plane. For the $\kappa$-deformed retarded prescription we take the
poles at
$\omega_{n}=\pm\omega_{\bf k} + {2\pi i n}/{q}-i\varepsilon$ ($n \in {\mathbb Z}$).
For convenience, the poles originally outside the real axis have
also been displaced by $-i\varepsilon$, which is of no consequence
due to the usual limit $\varepsilon\rightarrow 0$ taken at the end
of the calculations. 
We replace the symbol $(\gamma)$ by $(+)$ or $(-)$ in (\ref{integralfourierespacialtemporal})
and (\ref{IntegraltemporalICalC}) to indicate that 
the $\kappa$-deformed retarded or advanced prescription has been enforced 
on the Green function. 
Now, by writing the hyperbolic sine in
(\ref{IntegraltemporalICalC}) in terms  of exponentials and changing
the integration variable to $z=2q\omega$, we obtain in the $\kappa$-retarded case:
\begin{widetext}
\begin{equation}\label{IntegraltemporalIaberta}
{\cal G}^{(+)}({\bf k},t-t^{\prime})= 
\frac{q}{\sinh(2q\omega_{\bf k})} \int_{-\infty}^{\infty} \frac{dz}{2\pi} \; e^{-i\frac{(t -t
^{\prime})}{2q}z}\frac{d}{dz} \log\left[e^z-e^{-2q\left(\omega_{\bf k}+i\varepsilon\right)}\right] \; -\; \Bigl\{\;i\epsilon
\longrightarrow -i\epsilon\;\Bigr\} \;.
\end{equation}
\end{widetext}
These integrals are in a form ready for an application of Cauchy's
integral \cite{Whittaker},  also known as the Argument Principle.
For $t -t ^{\prime}>0$ the integrations on the real axis in
the two integrals (\ref{IntegraltemporalIaberta}) are extended to a
closed contour with the usual infinite-radius semicircle on the
lower imaginary half-plane, while for $t -t ^{\prime}<0$, the
contour is closed by an infinite-radius semicircle on the upper
imaginary half-plane. A straightforward calculation gives us for the
Green function (\ref{IntegraltemporalIaberta}) in the reciprocal
space the expression
\begin{widetext}
\begin{equation}\label{resultadofinalpolopbaixo}
{\cal G}^{(+)}({\bf k},t-t^{\prime}) = -
q\,\frac{\sin\left[\omega_{\bf k}(t -t
^{\prime})\right]}{\sinh(2q\omega_{\bf k})} \left\{\frac{\Theta(t
-t ^{\prime})}{1-\mbox{e}^{-\frac{2\pi (t -t ^{\prime})}{q}}}-
\Theta(t ^{\prime}-t )\frac{1}{2}\left[\coth\left(\frac{\pi(t
^{\prime}-t )}{q}\right)-1\right]\right\}.
\end{equation}
\end{widetext}
Substituting (\ref{resultadofinalpolopbaixo}) in
(\ref{integralfourierespacialtemporal}) and performing the angular
integral, we obtain
\begin{widetext}
\vspace{-0.3cm}
\begin{eqnarray}\label{funcaoGreenEspacialdk}
G^{(+)}(x,x^{\prime})\!&=&\!\frac{-1}{4\pi} \left\{ \frac{\Theta(t-t
^{\prime})}{1-\mbox{e}^{-\frac{2\pi (t -t^{\prime})}{q}}} +\Theta(t
^{\prime}-t)\frac{1}{2}\left[\coth\left(\frac{\pi(t ^{\prime}-t
)}{q}\right)-1\right] \right\}
\nonumber \\
&&\hspace{-17pt} \times
\frac{1}{|{\bf x}-{\bf x}^{\prime}|}
\int_0^{\infty}\!\!\frac{2\,dk}{\pi\sqrt{1+q^2k^2}}
\sin\left[\frac{|t -t ^{\prime}|}{q}\sinh^{-1}(qk)\right]
\sin\left(k|{\bf x}-{\bf x}^{\prime}|\right) \; ,
\end{eqnarray}
\end{widetext}
which is the $\kappa$-retarded Green function in space-time in terms of
a single quadrature. In the limit $q\rightarrow 0$ of no deformation
the term containing $\Theta(t^{\prime}-t )$ goes to zero,
the coefficient of $\Theta(t-t^{\prime})$ reduces to $1$
and the integral tends to a Dirac delta function. In this way we obtain from
(\ref{funcaoGreenEspacialdk}) the correct non-deformed limit
$ \left.G^{(+)}(x,x^{\prime})\right|_{q=0} =
\mbox{\large$\frac{-\Theta(t-t ^{\prime})}{4\pi|{\bf x}-{\bf x}^{\prime}|}$}
\;\delta\left[t^{\prime}-\left(t-|{\bf x}-{\bf
x}^{\prime}| \right)\right]$.
The $\kappa$-retarded Green function (\ref{funcaoGreenEspacialdk}) also
exhibits the zero limit for the remote past, namely,
$G^{(+)}(x,x^{\prime})=0$ for $t \rightarrow -\infty$, as in the non-deformed case.
However, in contrast with the non-deformed case, the $\kappa$-retarded Green function
(\ref{funcaoGreenEspacialdk}) has also contribution from the future,
stemming from the term containing $\Theta(t ^{\prime}-t )$ in
(\ref{funcaoGreenEspacialdk}). It is interesting to notice that the
existence of such a contribution from the future is natural if we
consider the deformed dispersion relation which leads to
(\ref{equacaogreen}) as describing phenomena at a fundamental
space-time scale $q$, at which the concepts of point and instant
lose their significance and should be substituted by the
concept of diffuse regions of space-time with dimensions of the
order of $q$. At this scale it is nothing but natural that
distinction between past and future loses its meaning
in a time interval of the order of $q$, thereby
giving the reason for the presence of the term containing
$\Theta(t^{\prime}-t )$ in the $\kappa$-retarded Green function
(\ref{funcaoGreenEspacialdk}). However, this interpretation must
pass the consistency checking that this contribution from the future
must die out rapidly when the measurement time is much greater than
the fundamental length $q$, {\it i.e.}, when $t ^{\prime}-t \gg q$.
This is by far satisfied by the $\kappa$-retarded Green function
(\ref{funcaoGreenEspacialdk}), due to the property
\begin{eqnarray}\label{Gret(t'-t>>q)}
\left.G^{(+)}({\bf x},t;{\bf
x}^{\prime},t^{\prime})\right|_{t^{\prime}-t\gg q} \sim
e^{-2\pi(t^{\prime}-t)/q}\sim 0 \; .
\end{eqnarray}
Expression (\ref{funcaoGreenEspacialdk}) for the $\kappa$-retarded
Green function with these properties, together with the analogous
results for the $\kappa$-advanced Green function, are the main result
of this letter.
We have seen that the $\kappa$-retarded Green function favors the
idea that at a fundamental time scale $q$ the concept of instant of
time should be substituted by the the concept of a fuzzy time
interval of the order of $q$. It is also interesting to ask if the
$\kappa$-deformed Green functions describe some sort of fuzziness in
space directions. In the Green functions this information appears as
the effect of the deformation upon the usual light cone described by
the Dirac delta function.
To address this question it is convenient to write the
$\kappa$-retarded Green function (\ref{funcaoGreenEspacialdk}) as
\begin{widetext}
\vspace{-0.4cm}
\begin{equation}\label{deltaq}
G^{(+)}(x,x^{\prime})=\frac{-\Theta_q(t-t ^{\prime})}{4\pi|{\bf
x}-{\bf x}^{\prime}|} \delta_q(x-x^{\prime})\;\;;\;\;\;\;\;
\delta_q(x-x^{\prime})=
\int_0^{\infty}\!\!\frac{2\,dk}{\pi\sqrt{1+q^2k^2}}
\sin\left[\frac{|t -t ^{\prime}|}{q}\sinh^{-1}(qk)\right]
\sin\left(k|{\bf x}-{\bf x}^{\prime}|\right) \; ,
\end{equation}
\end{widetext}
where we have defined $\Theta_q(t-t ^{\prime})$ as the combination
of Heaviside functions between braces in
(\ref{funcaoGreenEspacialdk}) and $\delta_q(x-x^{\prime})$
as the integral in (\ref{funcaoGreenEspacialdk}). Since
$\lim_{q\rightarrow 0}\delta_q(x-x^{\prime})=
\delta\left(|t-t^{\prime}|-|{\bf x}-{\bf x}^{\prime}|\right])$, the
integral $\delta_q(x-x^{\prime})$ defined in (\ref{deltaq}) itself
is a smeared Dirac delta function representing a fuzzy light cone in
the deformed case. Naturally, the smaller is $q$ compared to
$|t-t^{\prime}|$ and $|{\bf x}-{\bf x}^{\prime}|$, the sharper is
the deformed light cone. The fuzzy light cone is clearly illustrated
in Fig. 1 as the plot of the integral $\delta_q(x-x^{\prime})$ in
(\ref{deltaq}) (for ${\bf x}={\bf 0}$, $t=0$,
$x^{\prime 2} = x^{\prime 3} = 0$ and $x^{\prime 1} = x^{\prime}$). 
Since $q=0.001$ AU in this figure, the
smeared light cone shows a definitely regular appearance for $x$ and
$t$ greater than $0.5$ AU. In Fig. 2 a contour plot depicts a light
cone leg in a range of $6000q$ with constant width and height except
for small oscillations to be expected from the nature of the
integrand in (\ref{deltaq}).

%%%%%%%%%%%%%%%%%%%%%%%%%%%%% Figure 1  %%%%%%%%%%%%%%%%%%%%%%%%
%
\begin{figure}[h!]
\centering
\includegraphics[width=6.5cm]{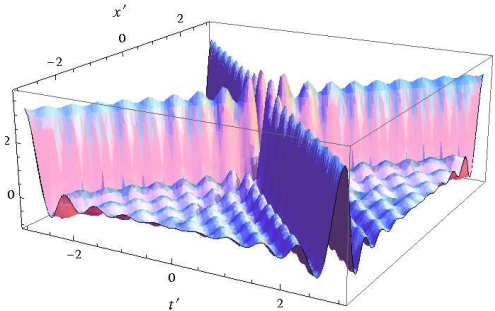}
\caption{A plot of $\delta_q$ given in (\ref{deltaq})
with $q=0.001$ AU.} 
\label{fig:figure1}
\end{figure}
%
%%%%%%%%%%%%%%%%%%%%%%%%%%%%% Figure 2  %%%%%%%%%%%%%%%%%%%%%%%%
%
\begin{figure}
\centering
\includegraphics[width=5.0cm]{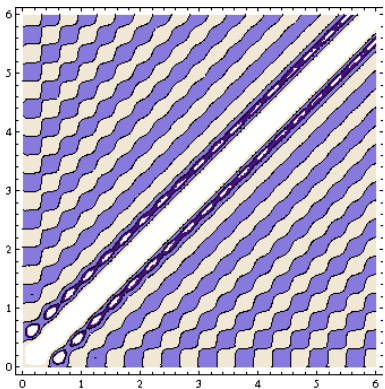}
\caption{A contour plot with contours at -0.5, 0 and 3.0 AU (darker
at lower levels)  of a light cone leg.} 
\label{fig:figure2}
\end{figure}

For the $\kappa$-retarded Green function the smeared
future light cone is exponentially suppressed according to our
fundamental result (\ref{Gret(t'-t>>q)}), while the smeared past
light cone is multiplied by the coefficient of $\Theta(t - t ^{\prime}
)$ in (\ref{funcaoGreenEspacialdk}), which tends to $1$ for
$t - t^{\prime}\gg q$. The little bumps outside the smeared light cone
and this smearing itself describe the Lorentz violation induced by
the deformation, which are negligible in the regime in which $q$ is
considered very small. Summing up, the $\kappa$-deformation smears
the light cone in spacetime and provides for the $\kappa$-retarded
Green function contributions from the past and an expected
contribution from an interval of future of the order of $q$. It is
interesting to compare the smearing of the light cone due to
$\kappa$-deformation with the change from the light cone to the so
called light wedge occurring in non commutative space
\cite{Alvarez-GaumeVazquez-Mozo2003,ChuFurutaInami2006}.

A completely analogous calculation leads, for the $\kappa$-advanced
Green function, to the result
$G^{(-)}(x,x^{\prime})=G^{(+)}(x^{\prime},x)$
with the correct non-deformed limit,
$\left.G^{(-)}(x,x^{\prime})\right|_{q=0}=\left.G^{(+)}(x^{\prime},x)\right|_{q=0}$,
and satisfying, in the distant future,
$\lim_{t \rightarrow \infty}G^{(-)}(x,x^{\prime})=0$.
In the $\kappa$-advanced Green function
the contributions from the blurred light cone, as expected, come
from both future and past. The contribution from the past comes from
an interval of time of the order of $q$  with an exponential decay
for measurement times much greater than this interval,
$ \left.G^{(-)}(x,x^{\prime})\right|_{t -t ^{\prime}\gg q}\! \sim
e^{-2\pi(t-t^{\prime})/q} \sim 0.$
Naturally, we can make comments on this property in analogy with the
ones made for the property (\ref{Gret(t'-t>>q)}) in the
$\kappa$-retarded case.

In this letter we presented  
the Green functions with the retarded and advanced prescription for the
$\kappa$-deformed dispersion relation. These
$\kappa$-retarded and $\kappa$-advanced Green functions have the
desired limits when the deformation goes to zero. Moreover, in the
case of small deformation, the deformed
Green functions exhibit a blurring in the distinction between past
and future inside time intervals of the order of the fundamental
length parameter $q$. In this way the $\kappa$ deformation appears
as an appropriate tool to describe the possibility of past and
future blurring at a fundamental length scale. This is a highly
desirable feature if we want to consider the deformation and its
length parameter $q$ as describing new physics at extremely small
scale, say at the Planck scale. We have also seen that the
$\kappa$-deformation also produces a blurring in the light cone.
It is rather surprising  that objects
such as those $\kappa$-deformed Green functions, which can be
defined from the sole data of a deformed dispersion relation,
exhibit the relaxed causality behavior to be expected in the realm
of QFT. It would be interesting to check if other deformed theories,
as for instance, the theory based on the $\kappa$-Poincar\'e algebra
in the bicrossproduct basis
\cite{MajidRuegg94,LukierskiRueggZakrewski95}, also exhibited the
nice {\lq\lq}relaxed causality{\rq\rq} properties described here.

\noindent
{\bf Acknowledgements}\\
The authors thank P.A. Maia Neto, Ioav Waga and R.R.R. Reis for
useful discussions. MJN and CF thank CNPq (brazilian agency) for
partial financial support.

\end{document}